\documentclass[pra, reprint]{revtex4-2}
\usepackage{amsmath, amssymb, mathtools, bm, bbold, booktabs, pgf} 
\usepackage[colorlinks=true,citecolor=blue,linkcolor=blue,urlcolor=blue, breaklinks]{hyperref}
\usepackage[mathscr]{eucal}
\begin{document}
\title{Integral Transforms in a Physics-Informed (Quantum) Neural Network setting: Applications \& Use-Cases}
\author{Niraj Kumar}
\affiliation{Pasqal SAS, 7 Rue Leonard de Vinci, 91300 Massy, France}
\author{Evan Philip}
\affiliation{Pasqal SAS, 7 Rue Leonard de Vinci, 91300 Massy, France}
\author{Vincent E. Elfving}
\affiliation{Pasqal SAS, 7 Rue Leonard de Vinci, 91300 Massy, France}
\date{\today}

\begin{abstract}
In many computational problems in engineering and science, function or model differentiation is essential, but also integration is needed. An important class of computational problems include so-called integro-differential equations which include both integrals and derivatives of a function. In another example, stochastic differential equations can be written in terms of a partial differential equation of a probability density function of the stochastic variable. To learn  characteristics of the stochastic variable based on the density function, specific integral transforms, namely moments, of the density function need to be calculated. Recently, the machine learning paradigm of Physics-Informed Neural Networks emerged with increasing popularity as a method to solve differential equations by leveraging automatic differentiation. In this work, we propose to augment the paradigm of Physics-Informed Neural Networks with \emph{automatic integration} in order to compute complex integral transforms on trained solutions, and to solve integro-differential equations where integrals are computed on-the-fly during training. Furthermore, we showcase the techniques in various application settings, numerically simulating quantum computer-based neural networks as well as classical neural networks.
\end{abstract}\maketitle

\section{Introduction}
Automatic differentiation allows computers to understand relationships between  inputs and outputs of functions and models. In particular, backpropagation~\cite{1986Rumelhart} allows to collect gradients of complex models, such as deep neural networks, in a highly efficient and exact manner. This is in contrast with approximate finite differencing methods.
Recently, the machine learning paradigm of Physics-Informed Neural Networks (PINNs~\cite{raissi2017physicsI, raissi2017physicsII, raissi2019physics, 2021Karniadakis}) emerged with increasing popularity as a method to solve, and to study systems governed by, differential equations. It relies on automatic differentiation in two aspects; differentiating with respect to model parameters to optimize and train the model, and differentiating with respect to input variables in order to represent derivatives of functions for solving differential equations~\cite{Yang2019, Meng2020, 2021Karniadakis}. Universal Function Approximators (UFAs) are used to represent complex solutions. Excellent candidates include classical  neural networks~\cite{raissi2017physicsI, raissi2017physicsII, raissi2019physics} and recently also quantum neural networks~\cite{Mitarai2018, Schuld2019a, kyriienko2021solving, Paine2021, Heim2021, Kyriienko2022} and quantum kernel functions~\cite{Havlicek2019, 2022Paine} were put forward. Representing differential equation solutions using universal function approximators allows to train flexibly~\cite{Raissi2018, Blechschmidt2021, Kollmannsberger2021, Lu2021}, but also gives access to accurate derivative expressions based on the trained solution; for example, after solving for the position of an oscillator as a function of time, one can compute the velocity or acceleration by automatic differentiation of the solution surrogate with respect to the time variable.

In this work, we propose to augment the PINN paradigm with automatic integration~\cite{lindell2021autoint}, in order to compute complex integral transforms of the solutions represented with UFAs, and to solve integro-differential equations~\cite{lakshmikantham1995theory} where integrals are computed on-the-fly during training. Furthermore, we showcase the techniques in various application settings, numerically simulating quantum computer-based neural networks as well as classical neural networks.

While prior work has shown methods to solve a range of different types of differential equations with (quantum) machine learning methods, including PDE~\cite{kyriienko2021solving} and SDE~\cite{2021Paine}, to the best of our knowledge we here show first how to solve integro-differential equations. Furthermore, we formalize automatic integration in the context of integral transforms, in contrast to prior work focusing on integration of just the function itself~\cite{lindell2021autoint}. Finally, we focus on settings where the machine learning model is trained based on data \emph{or} physics, i.e. differential equations and boundary conditions in the loss function, in contrast to prior work on automatic integration.

In Methods, we first briefly summarize earlier works on UFA, PINN, DQC, Automatic Integration, and Integral Transforms/kernels. Then, we show how these concepts can be combined into one workflow~\cite{patent}. In Results, we showcase various examples including calculation of moments in option pricing, moment of inertia, and electric potential due to complex charge densities.

\section{Methods}
\subsection{Physics-Informed (Quantum) Neural Networks}

Differential equations are ubiquitous in fields that utilise quantitative analysis. In general, solving differential equations is not easy and analytic solutions exist only in very specific cases, so approximate solution techniques are of great interest. The idea of using a ``function approximator" to represent the approximate solution of a general class of differential equations can be seen even in early techniques, such as the power series employed in the Frobenius Method~\cite{Frobenius1873}. Advancements in machine learning combined with the Universal Approximation Theorem~\cite{Hornik1989MultilayerFN} has made it possible to approximate functions using artificial neural networks. This enabled the development of physics-informed machine learning~\cite{2021Karniadakis} and novel techniques for solving differential equations representing the approximate solution using an artificial neural network as the universal function approximator (UFA). The use of physics-informed neural networks (PINNs)~\cite{raissi2019physics, raissi2017physicsI, raissi2017physicsII} to solve partial differential equations (PDEs) offers more flexibility (for example, incorporating data) and generality (for example, high-dimensional~\cite{2014Bach,2019Bauer} parameterized PDEs) compared to classical techniques. PINNs also led to the development of quantum-ready algorithms to solve PDEs, employing a quantum neural network, or `differentiable quantum circuit' (DQC)~\cite{kyriienko2021solving} as the UFA, offering an expanded latent space for representing expressive functions.

Physics informed machine learning proposes to solve PDEs by representing the approximate solution by an artificial neural network, differentiable quantum circuit (DQC) or another UFA and optimising the UFA with respect to a loss function constructed from the PDEs. A parameterised PDE can be formally represented by 
\begin{equation}
    \mathcal{P} (f, x; \lambda) = 0, 
\end{equation}
where $\mathcal{P}$ is a nonlinear operator acting on $f(x)$ and $\lambda$ is the parameter (note that $x$ and $\lambda$ could be multidimensional). For example, in case of the PDE
\begin{equation}
    \alpha\, \frac{\partial}{\partial p}g(p,q) + \beta\, \frac{\partial}{\partial q}g(p,q) + \gamma\, g(p,q) + \sin(q) = 0, 
\end{equation}
the parameters are $\alpha$, $\beta$, $\gamma$ and $\delta$ and $f(x)$ is $g(p,q)$. If the solution approximated by the UFA is represented by $u_\theta$, the convergence of  $u_\theta$ to the solution of the PDE is ensured (in most cases) by 
\begin{equation}
    \int \Vert \mathcal{P} (u_\theta, x; \lambda) \Vert = 0.
\end{equation}
The loss function to solve PDEs in physics informed machine learning is inspired by the above observation and is defined as
\begin{equation}
    \sum_{x_i} \Vert \mathcal{P} (u_\theta, x_i; \lambda) \Vert = 0,
\end{equation}
where $x_i$ represents a finite number of points chosen from the domain of the PDE. UFAs like artificial neural networks and differentiable quantum circuits (DQC) facilitate the easy evaluation of this loss function through automatic differentiation~\cite{1986Rumelhart,  Mitarai2018, Schuld2019a, Kyriienko2021}.  

While the above condition results in general solutions to the differential equation, boundary conditions (BCs) are needed to specify specific solutions. BCs can be included in the same fashion since they can also be formally represented by some nonlinear operator acting on a subset of the domain of the PDE and the two losses can be added. It is also possible to impose restrictions on the solution by choosing a UFA appropriately with inductive biases (for example, if we know that the solution passes through the origin, we can choose a UFA which always passes through the origin). For the interested reader, \textcite{2021Zubov} provides an updated technical overview of using PINNs to solve PDEs. 

\subsection{Automatic Integration}

Consider an artificial neural network 
\begin{equation}
\begin{aligned}
    f_\theta \vcentcolon \mathbb{R} \times \mathbb{R}^n &\rightarrow {\mathbb{R}}\\
    (s, \bm{v}) &\mapsto f_\theta (s, \bm{v}).
\end{aligned}
\end{equation}
By the fundamental theorem of calculus, the partial antiderivative $F_\theta$ of $f_\theta$ obeys
\begin{equation}\label{eq:ft_calc}
f_\theta (s, \bm{v}) = \int^{s} \left.\frac{\partial f_\theta(q, \bm{v})}{\partial q} \right\rvert_{q=s'} \;ds' = \int^{s} F_\theta(s', \bm{v})\;ds'. 
\end{equation}

Notice that $F_\theta(s, \bm{v})$ is also a function with the same domain and codomain of $f_\theta(s, \bm{v})$ and can be represented by an artificial neural network which shares the same parameters $\theta$ with $f_\theta$. \textcite{lindell2021autoint} point out that one could optimise $F_\theta$ to represent some target output data and obtain $f_\theta$, without any additional optimisation, from the optimised parameters of $F_\theta$. Since this procedure obtains the neural network of the antiderivative ``automatically'' just like automatic differentiation obtains a neural network of the derivative, the term \emph{automatic integration} is coined. 

We propose to augment the PINN paradigm with automatic integration to compute the integral of PINN-learned functions; furthermore, we expand the application from function integration to the more general concept of \textit{integral transforms} on functions.

\subsection{Integral Transforms}
An integral transform is a mathematical operator which maps a function $f$ to another function $\hat{f}$ defined by   
\begin{equation}
    \hat{f}(\bm{x}, \bm{y}) = \int k(\bm{x}, \bm{x}')\, f(\bm{x}', \bm{y})\;d\bm{x}'. 
\end{equation}
The function $k(\bm{x}, \bm{x}')$ is known as the \emph{kernel} of the integral transform. 

Integral transform arises in a wide range of fields. It is often used to transform a problem that is too difficult to solve in its original representation, like transforming a differential equation to an algebraic equation via Fourier transform. It also arises frequently in fields utilising probability or statistics, for example, for Gaussian smoothing of data. 
 
If the kernel is identity, the integral transform reduces to an ordinary integral. There are various standard transforms, like the Fourier transform and Laplace transform, whose kernel is fixed. The kernel of fundamental solutions of PDEs and Green's functions, on the other hand, can depend on the PDEs and their boundary conditions.

In addition to adding simple automatic integration to the PINN paradigm, with this work, we propose to:
\begin{itemize}
    \item extend both data-driven and physics informed machine learning with a novel method to calculate integral transforms.
    \item extend the PINN paradigm with a method to calculate the integral transform of PINN-learned functions. 
    \item extend the PINN paradigm with a method to solve integro-differential equations.
    \item extend the above capabilities to quantum machine learning through differentiable quantum models like DQC~\cite{kyriienko2021solving} and QKR~\cite{2022Paine}. 
\end{itemize}

\subsection{Integral Transforms for PINN-learned Functions} \label{section:integral-transform}

In this work, we combine physics-informed (quantum) neural networks and automatic integration to formulate a quantum-compatible machine learning method of performing a general integral transform on a learned function. Consider an unknown function 
\begin{equation}
\begin{aligned}
    f \vcentcolon \mathbb{R} \times \mathbb{R}^n &\rightarrow {\mathbb{R}}\\
    (s, \bm{v}) &\mapsto f(s, \bm{v}).
\end{aligned}
\end{equation}
Some information about the function $f$ is known in terms of 
\begin{itemize}
    \item value of $f$ or its derivatives at certain parts of the domain.
    \item integro-differential equations of the unknown function $f$ in terms of variables $s$ and $\bm{v}$.
\end{itemize}
We discuss a method to express, in the PINN-setup, the integral transform
\begin{equation}
    \hat{f}(s', \bm{v}) = \int_{s_a(s')}^{s_b(s')} k(s, s')\, f(s, \bm{v})\;ds, 
\end{equation}
where the kernel function $k(s, s') \in \mathbb{R}$, $s_a(s') \in \mathbb{R}$, and $s_b(s') \in \mathbb{R}$ are known functions. This method is then used to either
\begin{itemize}
    \item estimate the integral transform of the function, without explicitly learning the function itself.
    \item solve the integro-differential equations for the unknown function $f$.
\end{itemize}

First, we define a surrogate function 
\begin{equation}
\begin{aligned}
    G \vcentcolon \mathbb{R} \times \mathbb{R} \times \mathbb{R}^n &\rightarrow {\mathbb{R}}\\
    (s, s', \bm{v}) &\mapsto G(s, s', \bm{v})
\end{aligned}
\end{equation}
in terms of the unknown function $f$ and known function $g$ as
\begin{equation}\label{eq:ufa_transform}
    \left. \frac{\partial G(q, s', \bm{v})}{\partial q} \right\rvert_{q=s} = k(s, s')\, f(s, \bm{v}).
\end{equation} 
If $s_a$ and $s_b$ are functions of $s'$, we define $k(s, s')=0$ if $s<s_a(s')$ or $s>s_b(s')$.
 
The next step is to recast the data and the (integro\discretionary{-)}{}{-)}differential equations that are given in terms of $f$ in terms of the surrogate function $G$. This is done by making the substitutions
\begin{align}\label{eq:transformDE}
    f(s, \bm{v}) &\rightarrow \frac{1}{k(s, s')} \left. \frac{\partial G(q, s', \bm{v})}{\partial q} \right\rvert_{q=s}\\
    \int_{s_a}^{s_b} k(s, s')\, f(s, \bm{v})\;ds &\rightarrow G(s_b, s', \bm{v}) - G(s_a, s', \bm{v}).
\end{align}

\begin{table}
\centering
\begin{tabular}{p{0.9\linewidth}}
\toprule
\multicolumn{1}{@{}l}{\textbf{1. Input}}\\
Combination of:\newline
(integro\discretionary{-)}{}{-)}differential equations, boundary conditions, data.\\
\midrule
\multicolumn{1}{@{}l}{\textbf{2. Transformation}}\\
Transform all input using:
\begin{equation*}
\begin{aligned}
f(s, \bm{v}) &\rightarrow \frac{1}{k(s, s')} \left. \frac{\partial G(q, s', \bm{v})}{\partial q} \right\rvert_{q=s}\\
\int_{s_a}^{s_b} k(s, s')\, f(s, \bm{v})\;ds &\rightarrow G(s_b, s', \bm{v}) - G(s_a, s', \bm{v}).
\end{aligned}
\end{equation*}
\\
\midrule
\multicolumn{1}{@{}l}{\textbf{3. Learning}}\\
Learn $G$ using a differentiable model (PINN, DQC, etc...).\\
\midrule
\multicolumn{1}{@{}l}{\textbf{4. Output}}\\
Use the learned $G$ to calculate the integral transform or the unknown function:
\begin{equation*}
\begin{aligned}
\int_{s_a}^{s_b} k(s, s')\, f(s, \bm{v})\;ds &= G(s_b, s', \bm{v}) - G(s_a, s', \bm{v})\\
f(s, \bm{v}) &= \frac{1}{k(s, s')} \left. \frac{\partial G(q, s', \bm{v})}{\partial q} \right\rvert_{q=s}.
\end{aligned}
\end{equation*}
\\
\bottomrule                          
\end{tabular}
\caption{Summary of the algorithm~\cite{patent}.
\label{fig:algo}}
\end{table}

With these substitutions, the entire problem is now stated in terms of $G$ and its derivatives, without any reference to $f$, its derivatives or its integral. The universal function approximator (UFA) in the PINN-setup, such as a classical neural network or DQC, can now be used to learn $G$ directly. Once learning is complete, the integral transform of the unknown function or the unknown function itself can be calculated from Eq.~\eqref{eq:transformDE}. A summary of the algorithm is shown in Table~\ref{fig:algo}.

\section{Results \& Discussion}
We next discuss several specific use-cases where integral transforms appear in a natural PINN setting.
\subsection{Quantitative finance - dealing with stochastic evolution \label{Sec:quant_fin}}

Stochastic evolution plays a pivotal role in quantitative finance. They emerge in computing trading strategies, risk, option/derivative pricing, market forecasting, currency exchange and stock price predictions, determining insurance policies among many other areas~\cite{leung2015optimal}. 

Stochastic evolution is governed by a general system of stochastic differential equations (SDEs) can be written as,
\begin{equation}
    dX_t = r(X_t, t)\, dt + \sigma(X_t, t)\, dW_t
    \label{SDE-1d}
\end{equation}
where $X_t$ is the stochastic variable parameterised by time $t$ (or other parameters). The deterministic functions $r$ and $\sigma$ are the \;drift and diffusion processes respectively. And $W_t$ is a stochastic Wiener process, also called the Brownian motion. The stochastic component introduced by the Brownian motion makes the SDEs distinct from other types of differential equations and resulting in them being non-differential and hence being difficult to treat directly. 

One way to treat the SDEs is to rewrite them in the form of a partial differential equation for the underlying probability density function function (PDF) $p(x, t)$ of the variables $x$ and $t$. This resulting equation is known as the Fokker-Planck equation or the forward Kolmogorov equation and can be derived using Ito's lemma, starting from the SDE equation~\cite{bogachev2022fokker},
\begin{equation}
    \frac{\partial p(x,t)}{\partial t} = -\frac{\partial}{\partial x}[r(x,t)\, p(x,t)] + \frac{1}{2}\frac{\partial^2}{\partial x^2}[\sigma^2(x,t)\, p(x,t)] 
\end{equation}

Solving the Fokker-Planck equation results in learning the PDF. The PDF itself does not provide insights into the statistical properties. In order to extract meaningful information from the distribution, the standard tool used is to compute the \emph{moments}~\cite{2015Alastair} which help describe the key properties of the distribution by computing the integral of a certain power of the random variable with it's PDF in the entire variable range. Essentially, it helps in building the estimation theory, a branch in statistics that provides numerical values of the unknown parameters on the basis of the measured empirical data that has a random component. Further it is also a backbone of the statistical hypothesis testing, an inference method used to decide whether the data at hand sufficiently supports a particular hypothesis~\cite{Hazelton2011}. The $m$-th order moment $I_m$ is defined as the integral transform of the PDF with the $m$-th power of $x$ as the kernel, that is
\begin{equation}
    I_m = \int_{x \in \mathcal{X}} x^m\, p(x,t) \;dx.
\end{equation}

In quantitative finance, the four most commonly used terms derived from moments are;
\begin{enumerate}
    \item mean, a first order moment which computes the expected return of a financial object
    \item variance, derived from second order moment which measures of spread of values of the object in the distribution and quantifies risk of the object
    \item skewness, derived from third order moment which  measures how asymmetric the object data is about it's mean
    \item kurtosis, derived from fourth order moment which focuses on the tails of the distribution and explains whether the distribution is flat or rather with a high peak
\end{enumerate}
 Higher order moments have also been considered to derive more intricate statistical properties of a given distribution~\cite{brillinger1991some}. 

In this section, we specifically focus on option pricing~\cite{smith1976option}. An option is a financial object which provides the holder with the right to buy or sell a specified quantity of an underlying asset at a fixed price (also called the strike price or exercise price) at or before the expiration date of the option. Since it is the right and not an obligation, the holder can choose not be exercise the right and allow the option to expire. There are two types of options:
\begin{enumerate}
    \item Call options: A call option gives the buyer of the option the right to buy the underlying asset at a fixed price, called the strike price, at any time prior to the expiration date. The buyer pays a price for this right. If at the end of the option, the value of asset is less than the strike price, the option is not exercised and expired worthless. If on the other hand, the value of the asset is greater than the strike price, the option is exercised and the buyer of the option buys the asset at the exercise price. And the difference between the asset value and the exercise price is the gross profit on the option investment. 
    \item Put options: A put option gives the buyer the right to sell the underlying asset at a fixed price, again called strike price or the exercise price at or prior to the expiration period. The buyer pays a price for this right. If the price of the underlying asset is greater than the strike price, the option is not exercised and is considered worthless. If on the other hand, the price of the asset is lower than the strike price, the option of the put option will exercise the option and sell the stock at the strike price. 
\end{enumerate}

The value of an option is determined by a number of variables relating to the underlying asset and financial markets such as current value of the underlying asset, variance in the value of the asset, strike price of the option, and time to expiration of the option. 

Let us first consider the call option. Let $X_t$ denote the price of the stock at time $t$. Consider a call option granting the buyer the right to buy the stock at a fixed price $K$ at a fixed time $T$ in the future, with the current time being $t = 0$. If at the time $T$, the stock price $X_T$ exceeds the strike price $K$, the holder exercises the option for a profit of $X_T - K$.  If on the other hand, $S_T \leq K$, the option expired worthless. This is the European option, meaning it can be exercised only at a fixed date $T$. An American or an Asian call option allows the holder to choose the time of exercise. The payoff of the option holder at the time $T$ is thus,
\begin{equation}
    (X_T - K)^{+} = \text{max}\{0, X_T - K\}
\end{equation}
To get the present value of this payoff, we multiply by a discount factor $e^{-rT}$, with $r$ a continuously compounded interest rate. Thus the present value of the option is,
\begin{equation}
    E[e^{-rT}(X_T - K)^{+}]
\end{equation}

For this expectation value to be meaningful, one needs to specify the distribution of the random variable $X_T$, the terminal stock price. In fact, rather than specifying the distribution at a fixed time, one introduces the model for the dynamics of the stock price. The general dynamics for a single random variable $X_t$ is given by the SDE in Eq~\ref{SDE-1d}.

In the simplest case corresponding to the European call/put option, both $r$ and $\sigma$ are constants and the payoff is computed at terminal time $T$  which leads to analytical closed form solution to the problem. However, for more complex Asian/American type options where the interest rate and volatility are current asset price dependent, and where the payoff is path dependent, there are no analytical solutions, thus leading to one resorting to Monte-Carlo techniques~\cite{boyle1977options}. 

We apply our integral transform method to compute the expected payoff for an option governed by the Ornstein-Uhlenbeck stochastic process to compute the expected payoff in the European and Asian call options. Ornstein-Uhlenbeck process describes the evolution of interest rates and bond prices, and in financial analysis its generalization is known as the Vasicek model~\cite{VASICEK1977177}. Ornstein-Uhlenbeck process also describes the dynamics of currency exchange rates, and is commonly used in Forex pair trading. Thus benchmarking option pricing with a SDE evolution governed by Ornstein-Uhlenbeck process produces an initial validation test-bed for our method. The Ornstein-Uhlenbeck process is governed by the SDE, 
\begin{equation}
    dX_t = \nu(\mu - X_t)\,dt + \sigma\, dW_t
\end{equation}
where the underlying parameters $(\nu, \mu, \sigma)$ are the speed of reversion $\nu$, long time mean level $\mu$, and the degree of volatility $\sigma$. Here we consider the process such that the speed of reversion $\nu > 0$, while the long term mean level $\mu = 0$. The corresponding PDE governing the evolution of the PDF function $p(x, t)$ for this process is called the Fokker-Planck equation and is given by,
\begin{equation}
    \frac{\partial p(x, t)}{\partial t} = \nu \frac{\partial}{\partial x}(x\,p) + \frac{\sigma^2}{2}\frac{\partial^2 p}{\partial x^2}
    \label{eq:Fokker-Planck-Ornstein-Uhlenbeck}
\end{equation}

\subsubsection{Computing expected payoff for European options \label{section:european-option}}
The European option expected payoff depends on the price of the option at the terminal time $T$. For simplicity, we consider the call option governed by the Ornstein-Uhlenbeck process for which the non-discounted expected payoff is $ E[(X_T - K)^{+}]$. The discounted payoff can similarly be computed with a slight additional computational overhead.  The expected value can be written as,
\begin{equation}
\begin{aligned}
     E[(X_T - K)^{+}] &= \left(E[X_T - K]\right)^{+} \\
     &= \left(\int_{x \in \mathcal{X}} x\,p(x,T)\;dx - K\right)^{+}
\end{aligned}
\end{equation}
where $p(x,T)$ is the PDF for the Ornstein-Uhlenbeck process at terminal time $T$. This process is simplistic enough to allow us to analytically derive the PDF valid for the Dirac delta initial distribution $p(x, t_\circ)$ peaked at $x_\circ$ that evolves as,
\begin{multline}    \label{eq:pdf-1d}
    p(x, t) = \\
    \sqrt{\frac{1}{\pi \times \mathscr{N}}}\exp\left[- \frac{(x - x_\circ \exp[-\nu(t - t_\circ)])^2}{\mathscr{N}} \right]
\end{multline}
where $\mathscr{N} = \frac{\sigma^2}{\nu} (1 - \exp[-2\nu(t - t_\circ)])$, with chosen parameters $\sigma = 1$, $\nu = 5$, $x_\circ = 2$, $t_\circ = 0$ and terminal time $T = 0.5$. 

In order to compute the expected payoff, we mimic the case where we first generate 50 independent samples of $x$ from the distribution $p(x,t)$ in the domain $[-5, 5]$ at time point $t = 0.1$. Subsequently we assume that the PDF is unknown and we are only provided the data samples. Further, we are given the PDE equation of the form Eq~\ref{eq:Fokker-Planck-Ornstein-Uhlenbeck}. Next, we use a $2 \times 10 \times 10 \times 1$ dense classical neural network, with hyperbolic tangent activation functions, to learn the $G$ function such that,
\begin{equation}
    \frac{\partial G}{\partial x} = x\,p(x, t)
    \label{eq:kappa-transfrom-mean}
\end{equation}
Further, we transform the PDE expressed in the form of $p(x,t)$ into a PDE written in the form of a $G$ function using Eq~\ref{eq:kappa-transfrom-mean} and \ref{eq:Fokker-Planck-Ornstein-Uhlenbeck},
\begin{equation}
    \frac{\partial^2 G}{\partial x \partial t} = \left(\nu \frac{\partial^2 G}{\partial x^2} + \frac{\sigma^2}{2}\frac{\partial^2}{\partial x^2}\left(\frac{1}{x}\frac{\partial G}{\partial x}\right) \right)x
    \label{eq:Ornstein-Uhlenbeck-Fokker-Planck-mean}
\end{equation}
For training, we first generate the $(x,t)$ grid of size (50, 20) where $x$ is linearly spaced from $[-5, 5]$ and $t$ is linearly spaced in the domain $[0.1, 0.5]$. At $t = 0.1$, we train our classical neural network using the Adam optimiser to minimise the loss function corresponding to the data-sample, while for the rest of the time points in the grid, we use the Eq~\ref{eq:Ornstein-Uhlenbeck-Fokker-Planck-mean} guidance. This is in accordance to the physics informed neural networks strategy~\cite{raissi2019physics, raissi2017physicsI, raissi2017physicsII}. 

\begin{figure}[!htb]
    \centering
    \input{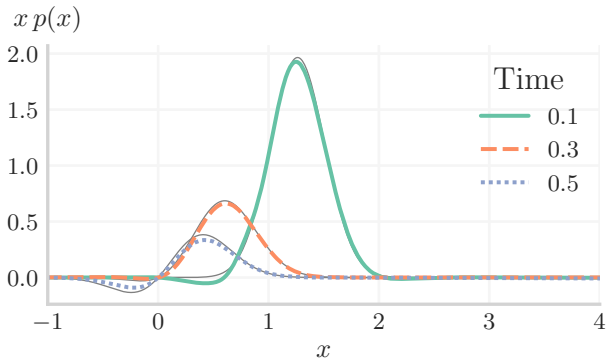}
    \caption{The thin grey lines show the theoretical value of $x \,p(x,t)$ in the $x, t$ domain. The colored lines show $\frac{\partial G}{\partial x}$. The matching of shapes indicate that the Neural network has learnt the correct function derivative.
}
    \label{fig:x-px-2.0}
\end{figure}
After training, we test our classical neural network performance by first comparing the shape of $\frac{\partial G}{\partial x}$ learnt by classical neural network vs the input shape of $x \,p(x,t)$ in Figure~\ref{fig:x-px-2.0}.
In order to provide a robust estimate of our expected payoff estimate, we also compute the standard deviation of payoff. The standard deviation is given by,
\begin{multline}
    E[X_T^2] - E[X_T]^2 =\\ 
    \int_{x \in \mathcal{X}} x^2\,p(x,T)\;dx - \left(\int_{x \in \mathcal{X}}x\,p(x,t)\;dx\right)^2 
\end{multline}

In order to estimate $E[X_T^2]$, we again use our integral transform method. We follow the same PDF sample generation process and subsequently estimating the expressional form of the PDF as highlighted before. Further, we consider the same size of the classical neural network  with the same training configuration as before to learn the function $G'$,
\begin{equation}
    \frac{\partial G'}{\partial x} = x^2\, p(x,t)
        \label{eq:zeta-transfrom-mean}
\end{equation}
Further, we transform the PDE expressed in the form of $p(x,t)$ into a PDE written in the form of a $G'$ function using Eq~\ref{eq:zeta-transfrom-mean} and \ref{eq:Fokker-Planck-Ornstein-Uhlenbeck},
\begin{equation}
    \frac{\partial^2 G'}{\partial x \partial t} = \left(\nu \frac{\partial}{\partial x} \left(\frac{1}{x}\frac{\partial G'}{\partial x}\right) + \frac{\sigma^2}{2}\frac{\partial}{\partial x^2}\left(\frac{1}{x^2}\frac{\partial G'}{\partial x}\right) \right)x^2
\end{equation}

\begin{figure}[!htb]
    \centering
    \input{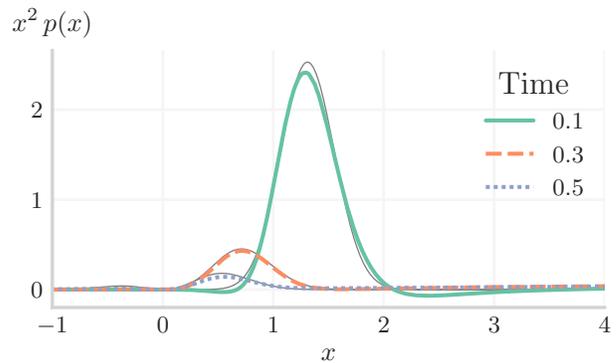}
    \caption{The thin grey lines show the theoretical value of $x^2 \,p(x,t)$ in the $x, t$ domain. The colored lines show $\frac{\partial G'}{\partial x}$. The matching of shapes indicate that the Neural network has learnt the correct function derivative.
}
    \label{fig:x^2-px-2.0}
\end{figure}
We compare the learnt shape of $\frac{\partial G'}{\partial x}$ with the input shape of $x^2 \,p(x,t)$ in Figure~\ref{fig:x^2-px-2.0} to showcase that the classical neural network has learnt the correct time-dependent shape. 

\begin{table}[!htb]
\begin{tabular}{c c} 
 \hline
Neural Network & Analytical\\ 
 \hline
 0.12$\pm$0.28 & 0.10$\pm$0.32 \\ 
\hline
\end{tabular}
\caption{\label{Table:1}}
\end{table}
Subsequently, for the strike price of $K = 0.06$ and terminal time $T = 0.5$, we estimated the expected payoff at the terminal time with two evaluations each of the learnt $G$ and $G'$ function, at points $x_\text{final} = 5$ and $x_\text{init} = -5$ and compare the classical neural network result with the analytical method for the Ornstein-Uhlenbeck PDF. We report the expected payoff value in Table~\ref{Table:1}.

\subsubsection{Computing expected payoff for Asian options}
Next we consider the Asian `path dependent' price option, where we in this example consider the dynamics of the underlying asset $X_t$ to be governed by the Ornstein-Uhlenbeck process. The Asian option requires simulating the paths of the asset not just at the terminal time $T$ , but also at intermediate times. For these options, the payoff depends on the average level of the underlying asset. This includes, for example, the non-discounted payoff $(\bar{X} - K)^{+} $ where,
\begin{equation}
    \bar{X} = \frac{1}{m}\sum_{j = 1}^m X_{t_j}
\end{equation}
for some fixed set of dates $0 = t_0 < t_1 \cdots < t_m = T$, with $T$ being the date where the payoff is received. The expected payoff is then $E[(\bar{X} - K)^{+} ]$. 

Since we have integrated the Fokker-Planck PDE from time $[0.1, 0.5]$ in Section~\ref{section:european-option}, this immediately provides us with the expected payoff and the standard deviation at each of the intermediate time points. With this, we can query the values at each time points with 2 evaluations of the trained model.

\subsubsection{Computing expected payoff for Basket options}
The above work focussed on SDEs and the corresponding Fokker-Planck equations with two dimensions. Here we look at an $N+1$ dimensional SDE and Fokker-Planck. In real world, an option depends on multiple underlying assets and hence strategies for a generalised Fokker-Planck equation solving is extremely relevant. These type of options are called \textbf{basket options} whose underlying is a weighted sum or average of different assets that have been grouped together in a basket~\cite{linders2014basket}. 

Consider a multivariate $N$ dimensional SDE system of $\bm{X_t} = (X_{1,t}, \cdots X_{N,t})$ with an M-dimensional Brownian motion $\bm{W_t} = (W_{1,t}, \cdots W_{M,t})$ given by,
\begin{equation}
    d\bm{X_t} = \bm{r}(\bm{X_t}, t)\, dt + \boldsymbol\sigma(\bm{X_t}, t)\, d\bm{W_t}
    \label{eq:generalised SDE}
\end{equation}

where $\bm{X_t}$ and $\bm{r}(\bm{X_t}, t)$ are $N$ dimensional vectors and $\boldsymbol\sigma(\bm{X_t}, t)$ is a $N \times M$ matrix.

In this specific case, we consider a basket option governed by two assets $X_1$ and $X_2$ where we only provided with the data samples $(x_1, x_2, t)$ from the underlying PDF $p(x_1, x_2, t)$. We consider that the underlying PDF corresponds to the independent Ornstein-Uhlenbeck processes of the with the following PDF,
\begin{multline}\label{eq:pdf-2d}
        p(x_1, x_2, t) = \\
        \sqrt{\frac{1}{\pi \times \mathscr{N}_1}}\exp\left[- \frac{(x_1 - x^\circ_1 \exp[-\nu_1(t - t^\circ_1)])^2}{\mathscr{N}_1} \right] \times \\
        \sqrt{\frac{1}{\pi \times \mathscr{N}_2}}\exp\left[- \frac{(x_2 - x^\circ_2 \exp[-\nu_2(t - t^\circ_2)])^2}{\mathscr{N}_2} \right],
\end{multline}
where $\mathscr{N}_i = \frac{\sigma_i^2}{\nu_i} \left(1 - \exp[-2\nu_1(t - t^\circ_i)]\right)$. We use $\sigma_1 = 1$, $\nu_1 = 5$, $x^\circ_1 = 2$, $t^\circ_1 = 0$, $\sigma_2 = 2$, $\nu_2 = 3$, $x^\circ_2 = 1$, and $t^\circ_2 = 0$.

The objective is to compute the expected payoff $E[\bar{X} - K]$ at terminal time $T$ where,
\begin{equation}
    \bar{X} = \frac{1}{2}(X_{1, T} + X_{2, T})
\end{equation}
Computing $E[\bar{X}]$ amounts to computing the maximum likelihood estimation,
\begin{equation}
    E[\bar{X}] = \int_{x_1}\int_{x_2} \frac{1}{2}(x_1 + x_2)\,p(x_1, x_2, T)\;dx_1 \;dx_2
\end{equation}
We first generate 50 independent samples each of $x_1$ and $x_2$ in the domain $[-5, 5]$ at each time point. Further we generate the data-samples at 15 different times points between $[0.3, 1]$. Subsequently we assume that the PDF is unknown and we are only provided the data samples. We then learn the histogram from the data and estimate the expressional form of the PDF function. 

Next, we use a $3 \times 10 \times 10 \times 10 \times 1$ dense classical neural network, with hyperbolic tangent activation functions,  to learn the function $G$ such that,
\begin{equation}
    \frac{\partial G}{\partial x_1 \partial x_2} = k(x_1, x_2)\,p(x_1, x_2, t)
\end{equation}
where $k(x_1, x_2) = \frac{1}{2}(x_1 + x_2)$.

\begin{figure}[!htb]
    \centering
    \input{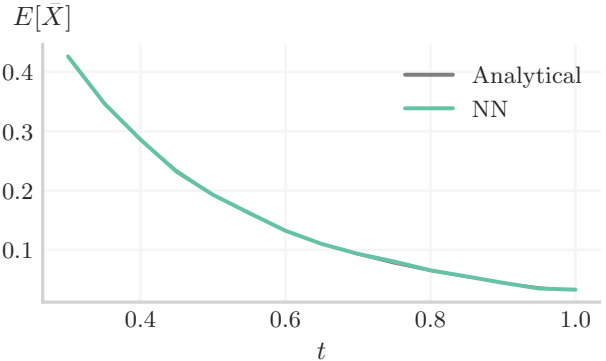}
    \caption{Plot of the maximum likelihood estimation $E[\bar{X}]$ with the classical neural network integral transform and analytical method over the time domain between [0.3, 1.0].}
    \label{fig:ou-mean-2d}
\end{figure}

We train our classical neural network with the Adam optimiser with a learning rate = 0.005 and a total of 20000 epochs. In order to validate our method, we estimate the maximum likelihood estimation at each time point with two evaluations of the learnt $G$ function, at points $(x_1, x_2)_\text{final} = (5,5)$ and $(x_1, x_2)_\text{init} = (-5, -5)$ and compare the classical neural network result with the analytical method. The results in Figure~\ref{fig:ou-mean-2d} validates the performance of our integral transform method for the specific case chosen.

\subsection{Calculation of Moment of Inertia}

Due to the ubiquity of applied statistics, integral transforms of the type discussed in Section~\ref{Sec:quant_fin}, namely moments, find applications in a wide range of fields. In addition, integral transforms of the same type also find applications in physics and engineering, distinct from those due to applied statistics. In fact, the mathematical concept was so named due to its similarity with the mechanical concept of moment in physics~\cite{walker1929}. The $n$-th moment of a physical quantity with density distribution $\rho(\bm{r})$, with respect to a point $\bm{r}_\circ$, is given by
\begin{equation}
    \mu_n(\bm{r}_\circ) = \int |\bm{r}-\bm{r}_\circ|^n\, \rho(\bm{r}) \;d^3r. 
\end{equation}
As an integral transform, the input function here is $\rho(\bm{r})$ and the kernel $k(\bm{r}, \bm{r}_\circ) = |\bm{r}-\bm{r}_\circ|^n$. Examples of physical quantities whose density could be represented by $\rho(\bm{r})$ are mechanical force, electric charge and mass. Here are some examples of moments in physics:
\begin{enumerate}
    \item Total mass is the zeroth moment of mass.
    \item Torque is the first moment of force.
    \item Electric quadruple moment is the second moment of electric charge.
\end{enumerate}

Here, we will look at the particular example of calculating the moment of inertia of a rotating body. Moment of inertia is the second moment of mass and the relevant moment of inertia for studying the mechanics of a body is usually the one about its axis of rotation. A spinning figure skater can reduce their moment of inertia by pulling their arm in and thus increase their speed of rotation, due to conservation of angular momentum. Moment of inertia is of utmost importance in engineering with applications ranging from design of flywheels to increasing the maneuverability of fighter aircrafts. 

\begin{figure}[!htb]
\begingroup%
  \makeatletter%
  \providecommand\color[2][]{%
    \errmessage{(Inkscape) Color is used for the text in Inkscape, but the package 'color.sty' is not loaded}%
    \renewcommand\color[2][]{}%
  }%
  \providecommand\transparent[1]{%
    \errmessage{(Inkscape) Transparency is used (non-zero) for the text in Inkscape, but the package 'transparent.sty' is not loaded}%
    \renewcommand\transparent[1]{}%
  }%
  \providecommand\rotatebox[2]{#2}%
  \newcommand*\fsize{\dimexpr\f@size pt\relax}%
  \newcommand*\lineheight[1]{\fontsize{\fsize}{#1\fsize}\selectfont}%
  \ifx\svgwidth\undefined%
    \setlength{\unitlength}{184.5bp}%
    \ifx\svgscale\undefined%
      \relax%
    \else%
      \setlength{\unitlength}{\unitlength * \real{\svgscale}}%
    \fi%
  \else%
    \setlength{\unitlength}{\svgwidth}%
  \fi%
  \global\let\svgwidth\undefined%
  \global\let\svgscale\undefined%
  \makeatother%
  \begin{picture}(1,0.64227642)%
    \lineheight{1}%
    \setlength\tabcolsep{0pt}%
    \put(0,0){\includegraphics[width=\unitlength,page=1]{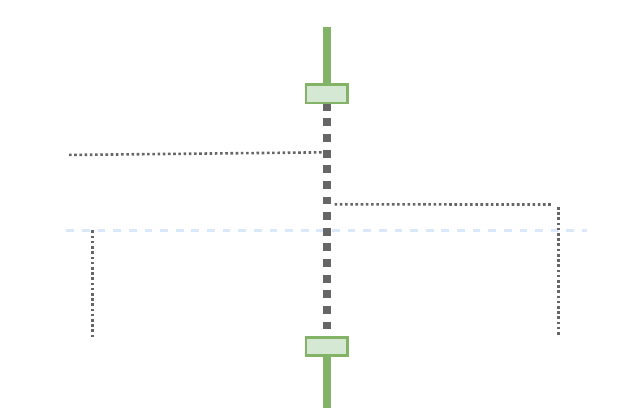}}%
    \put(0.30691057,0.42682927){\makebox(0,0)[t]{\lineheight{1.25}\smash{\begin{tabular}[t]{c}$R$\end{tabular}}}}%
    \put(0,0){\includegraphics[width=\unitlength,page=2]{Rotation.pdf}}%
    \put(0.18089431,0.19512195){\makebox(0,0)[t]{\lineheight{1.25}\smash{\begin{tabular}[t]{c}$h_\circ$\end{tabular}}}}%
    \put(0,0){\includegraphics[width=\unitlength,page=3]{Rotation.pdf}}%
    \put(0.63414634,0.53902439){\makebox(0,0)[t]{\lineheight{1.25}\smash{\begin{tabular}[t]{c}$\omega$\end{tabular}}}}%
    \put(0,0){\includegraphics[width=\unitlength,page=4]{Rotation.pdf}}%
    \put(0.77845528,0.35162602){\makebox(0,0)[t]{\lineheight{1.25}\smash{\begin{tabular}[t]{c}$\big(r,h(r)\big)$\end{tabular}}}}%
    \put(0,0){\includegraphics[width=\unitlength,page=5]{Rotation.pdf}}%
  \end{picture}%
\endgroup%

    \caption{Spinning vessel containing an incompressible liquid. At rest, the fluid reaches a height $h_0$, and when spinning at equilibrium, the height of the liquid depends on the distance to the center of the vessel.}
    \label{fig:Moment-of-Inertia-Expt}
\end{figure}
For our example, we consider a rectangular vessel containing a liquid of mass density $\rho$ that is rotated at an angular frequency $\omega$ about one of its face-centered axes, as shown in Figure~\ref{fig:Moment-of-Inertia-Expt}. When the vessel is at rest, the liquid is initially at a uniform height $h_\circ$. As the frequency of rotation $\omega$ is increased, the shape of the surface of the liquid changes due to centrifugal force and the distribution of mass changes. This causes the moment of inertia to change as well. For simplicity, we will consider an ideal incompressible fluid ($\rho$ is a fixed constant) and the width of the vessel (perpendicular to radial direction) to be negligible, i.e. $w\ll R$. 

The equilibrium height $h(r)$ of the surface of the liquid at radial distance $r$ from the axis of rotation at frequency of rotation $\omega$ is given by Bernoulli's principle, which is represented in differential form as
\begin{equation}\label{eq:bernoulli}
    \frac{\partial}{\partial r}(\frac{1}{2} \rho r^2 \omega^2 + \rho g h(r) + p_\text{air}) = 0,
\end{equation}
where $\rho$ is the density of the liquid, $g$ is the acceleration due to gravity and $p_\text{air}$ is the air pressure.  

The moment of inertia of this system about its axis of rotation is given by
\begin{equation}\label{eq:moi}
    I = \int_{0}^{R} r^2 \rho h(r) w \;dr = \rho w \int_{0}^{R} r^2 h(r) \;dr, 
\end{equation}
where $R$ is the radius of the container and $w$ is the width. $\rho h(r) w \;dr$ is the total mass at distance $r$ and we are calculating the second moment of mass. Conservation of total fluid volume fixes the constant of integration.

Assuming the situation is not simple enough to solve analytically, conventionally one first solves Eq.~\eqref{eq:bernoulli} for $h(r)$ and then proceed to perform the integral Eq.~\eqref{eq:moi} numerically. In contrast, by combining PINN and automatic integration, $I$ is calculated directly, without first having to find $h(r)$. To accomplish
 this, we first transform $h(r)$ in Eq.~\eqref{eq:bernoulli} according to Eq.~\eqref{eq:ufa_transform} using 
 \begin{equation}
   \frac{\partial G(r)}{\partial x} = r^2 h(r)
 \end{equation}
 resulting in 
\begin{equation}\label{eq:bernoulli_transformed}
    \frac{\partial}{\partial r}\left(\frac{1}{2} \rho r^2 \omega^2 + \rho g \left(\frac{1}{r^2}\frac{\partial G(r)}{\partial x}\right) + p_\text{air}\right) = 0. 
\end{equation}
Now, the moment of inertia $I$ is given by
\begin{equation}\label{eq:moi_transformed}
    I = \rho w (G(R) - G(0)). 
\end{equation}

\begin{figure}[!htb]
    \centering
    \input{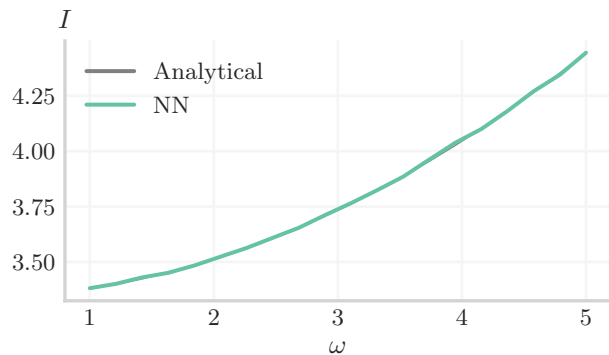}
    \caption{Moment of inertia $I$ as a function of vessel spinning frequency $\omega$. We compare the machine-learned value with the analytical solution and find a close correspondence for all values of $\omega$.}
    \label{fig:Moment-of-Inertia-Result}
\end{figure}

We use a dense classical neural network to calculate the surrogate function $G$ according to the physics informed neural networks strategy~\cite{raissi2019physics, raissi2017physicsI, raissi2017physicsII}, where the loss terms are constructed according to the PDE in Eq.~\eqref{eq:bernoulli_transformed}. Upon convergence, the moment of inertia is calculated according to Eq.~\eqref{eq:moi_transformed} and the result is plotted along with the analytical solution for various values of $\omega$ in Figure~\ref{fig:Moment-of-Inertia-Result}. Note that we have not calculated the intermediate quantity $h(r)$ with the classical neural network to obtain this final result, and only two queries to the anti-derivative neural network were required per configuration $\omega$.

\subsection{Electric Potential of Dynamic Charge Density}

The integral transforms so far have used kernels of the form of a positive power of $x$. However, the applications of integral transform is very broad and encapsulate techniques like Fourier transform, Green's functions, fundamental solutions, and so on which rely on different kernel functions. There are various standard transforms like Fourier transform whose kernel does not depend on the problem at hand; the kernel of fundamental solutions of PDEs and Green's functions, on the other hand, can depend on the PDEs and their boundary conditions. Though covering all types of integral transforms will not be practical, in order to illustrate the broad applicability of this technique, we will demonstrate the use of another type of kernel function. 

The Laplace equation is a PDE that arises in physical laws governing phenomenon ranging from heat transfer to gravitational force. Here, we find the electric potential in free space due to a distribution of electrical charge evolving in time according to the advection-diffusion equation, using the Laplace equation. We will calculate the electric potential through integral transform using the fundamental solution of the 3D Laplace equation as the kernel
\begin{equation}
    k(\bm{r}, \bm{r}_\circ) = \frac{1}{|\bm{r}-\bm{r}_\circ|},
\end{equation}
without explicitly calculating the charge distribution using the advection-diffusion equation. Note that `3D' refers to the dimensions of space here, which differs from the number of dimensions we integrate over. The number of dimensions we integrate over, in the limit of a thin tube, is 1.

\begin{figure}[!htb]
    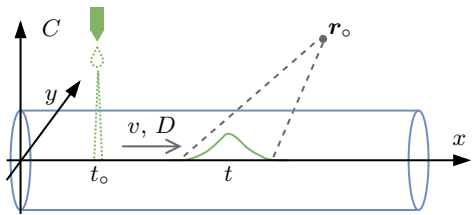
    \caption{Situational sketch of electric potential as observed at a position $\bm{r}_\circ$ relative to a tube of flowing liquid, into which a drop of charged particles is dropped which subsequently drift and diffuse in the liquid.}
    \label{fig:Potential-Expt}
\end{figure}

Consider a \;drop of charged ions \;dropped into a tube of (neutral)\, flowing liquid, as shown in Figure~\ref{fig:Potential-Expt}. For simplicity, we consider the flow to be one-dimensional. The ions \;drift and diffuse with time as governed by the advection-diffusion equation, moving forward with velocity $v$ and diffusing with diffusion coefficient $D$. Our goal is to find the electric potential at an arbitrary point $(x_\circ, y_\circ, z_\circ)$ due to the ions, as a function of time.

The concentration field $C(x, t)$ of ions in one-dimensional flow along $y = z = 0$ at point $x$ at time $t$ is given by the advection-diffusion equation
\begin{equation}\label{eq:ad-diff}
\frac{\partial C}{\partial t} +v \frac{\partial C}{\partial x}
 = D \frac{\partial^2 C}{\partial x^2},
\end{equation}
where $v$ is the fluid velocity and $D$ is the diffusion constant.

The potential at a point $\bm{r}_\circ = (x_\circ, y_\circ)$ is given by the integral 
\begin{equation}\label{eq:potential}
    V(\bm{r}_\circ , t) = \lambda \int_{- \infty}^{ \infty} \frac{1}{|\bm{r}-\bm{r}_\circ|} C(x,t) \;dx, 
\end{equation} 
where $\lambda$ is simply a constant of proportionality and $\bm{r} = (x, 0, 0)$.

As before, we first transform $C(x, t)$ in Eq.~\eqref{eq:ad-diff} according to Eq.~\eqref{eq:ufa_transform} using
\begin{equation}
    G(\bm{r},\bm{r}_\circ, t) = \frac{1}{|\bm{r}-\bm{r}_\circ|} C(x,t)
\end{equation}
resulting in 
\begin{equation}\label{eq:ad-diff_transformed}
|\bm{r}-\bm{r}_\circ| \frac{\partial G}{\partial t} + v \frac{\partial}{\partial x}\left(|\bm{r}-\bm{r}_\circ| G\right)
 = D \frac{\partial^2}{\partial x^2}\left(|\bm{r}-\bm{r}_\circ| G\right).
\end{equation}
The potential is now given by
\begin{equation}\label{eq:potential_transformed}
    V(\bm{r}_\circ , t) = \lambda (G(\infty, \bm{r}_\circ, t) - G(-\infty, \bm{r}_\circ, t)). 
\end{equation}
In practice, we choose finite limits that are sufficiently large such that the contributions from outside these limits is negligible.

\begin{figure}[!htb]
    \centering
    \input{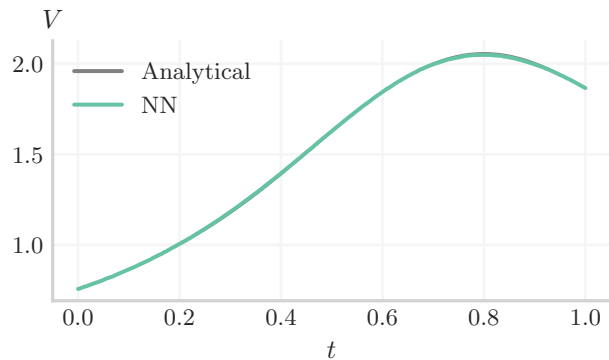}
    \caption{Plotting the electric potential computed as the potential integral transform over the charge concentration as depicted in the situational sketch Figure~\ref{fig:Potential-Expt}.}
    \label{fig:Potential-Result}
\end{figure}
We use a dense classical neural network to calculate the surrogate function $G$ according to the physics informed neural networks strategy~\cite{raissi2019physics, raissi2017physicsI, raissi2017physicsII}, where the loss terms are constructed according to the PDE in Eq.~\eqref{eq:ad-diff_transformed}. Upon convergence, the potential is calculated according to Eq.~\eqref{eq:potential_transformed} by selecting two extreme points beyond which the change in $G$ is sufficiently small. The result is plotted along with the analytical solution in Figure~\ref{fig:Potential-Result}. Note that we have not calculated the intermediate quantity $C(x, t)$ with the classical neural network to obtain this final result.

\subsection{Population Growth}
Lastly, we we use a simulated DQC to solve a simple integro-differential equation. We use an inhomogeneous Volterra–Fredholm integro-differential equation~\cite{2011Baykus}, inspired by the evolution of female population~\cite{2020Pratap}, given by
\begin{equation}\label{eq:VF}
    \frac{\partial}{\partial t} b(t) = u(t) + \int_0^{t} k(t, t')\, b(t') \;dt', 
\end{equation}
where $b(t)$ is the number of female births at time $t$, $k$ is the net maternity function and $u$ is the rate of female births due to the population already present at $t=0$. We take $k(t, t') = t - t'$, $u(t) = (6(1+t) - 7 \exp(t/2) - 4 \sin (t))/4$, and the initial condition $b(0) = 1$ and attempt to solve this. In this particular case, we can solve the equation exactly and the solution is $b(t) = (\exp(t/2)-\sin(t)+\cos(t))/2$

First, we apply Leibniz integral rule to take the total derivative of Eq.~\eqref{eq:VF}, which yields
\begin{equation}
    \frac{\partial^2}{\partial t^2} b(t) = \frac{\partial}{\partial t}u(t) + \int_0^{t} b(t') \;dt',
\end{equation}
since the contribution from the boundary term disappears. We do this to avoid the singularity that would otherwise arise from the inverse of the original kernel in the integral. Now, we transform the resulting equation according to Eq.~\eqref{eq:transformDE} with 
\begin{equation}
    \frac{\partial}{\partial t} G(t) = b(t)
\end{equation}
to obtain 
\begin{equation}\label{eq:VT_transformed}
    \frac{\partial^3}{\partial t^3} G(t) = \frac{\partial}{\partial t}u(t) + G(t) - G(0).
\end{equation}
As per the original boundary condition, 
\begin{equation}
    \left. \frac{\partial}{\partial t} G(t) \right\rvert_{t=0} = 1.
\end{equation}
In addition, by setting $t=0$ in Eq.~\eqref{eq:VF}, we get an additional boundary condition 
\begin{equation}
    \left. \frac{\partial^2}{\partial t^2} G(t) \right\rvert_{t=0} = m(0) = -1/4. 
\end{equation}
We set $G(0) = 0$ by choice. 

\begin{figure}[!htb]
    \centering
    \input{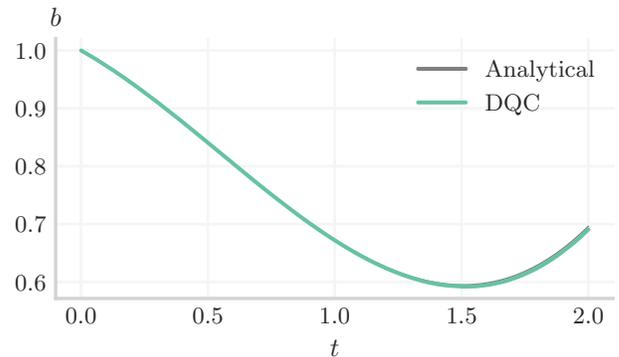}
    \caption{Solution to the population growth integro-differential equation written in \eqref{eq:VF}. The female population $b(t)$ is plotted as a function of time, comparing our quantum neural network automatic integration approach to the analytical solution.}
    \label{fig:integrodifferential}
\end{figure}
We calculate the surrogate function $G(t)$ according to the physics informed neural networks strategy~\cite{raissi2019physics, raissi2017physicsI, raissi2017physicsII}, where the loss terms are constructed according to the PDE in Eq.~\eqref{eq:VT_transformed}. As the UFA, we use a 4 qubit DQC, with a Chebyshev map~\cite{kyriienko2021solving, raissi2017physicsI, raissi2017physicsII} as the feature map, depth 4 Hardware Efficient Ansatz~\cite{Kandala2017} as the variational circuit and total magnetization as the cost function. The boundary conditions were enforced using exponential pinning of the UFA and the optimization was performed at 25 uniformly spaced points in the domain using AdaBelief~\cite{AdaBelief}. The derivative of $G(t)$ learned in 411 epochs is plotted along with the analytical solution for the unknown function $b(t)$ in Figure~\ref{fig:integrodifferential}.

\section{Conclusion}

In this work, we showcased how automatic integration can be used in a physics-informed machine learning setting, using either classical neural networks or quantum models such differentiable quantum circuits. This method can, however, be extended other models, the key requirement being the ability of the model to be differentiated accurately and efficiently with respect to its input as well as its model parameters. Quantum kernel functions~\cite{Havlicek2019, 2022Paine} or hybrid quantum-classical models are of particular interest. 

In \textcite{2022Varsamopoulos} it is shown how quantum machine learning models can be trained and subsequently its solutions extremized, with applications in design optimization and beyond. Of particular interest would be to combine such approach with the approach from the current work. While solutions can be extremized at certain points, integrals over such solutions may be extremized as well to satisfy some design requirement such as average value tolerances, total flux limits, etc.

In \textcite{Heim2021, Both2021} it is shown how (quantum) machine learning models can learn, or discover, differential equations based on observations (data) on the target function. It is a target of future research to combine the current work with model discovery in two directions: fitting integro-differential equation models to data, or performing integral transforms on the solution functions in a model discovery setting. Both of these cases can be done either with known terms and unknown coefficients, or given a library of candidate terms and a regularization term.

Interesting integral transforms for future studies in this context include Fourier transform, Laplace transform as well as field integrals and other non-local operators. While Eq.~\eqref{eq:transformDE} looks suspiciously simple, the transformation can be complicated by boundaries in higher dimensions or singularities. Techniques such as regularisation may be explored in the context of physics informed machine learning to tackle these problems.

We leave for future work a more rigorous comparison of end-to-end run-time costs between this automatic integration approach and numerical/statistical integration methods. 


\emph{Disclosure:} A patent application for the method described in this manuscript has been submitted by PASQAL SAS with NK, EP and VEE as inventors~\cite{patent}.

\bibliography{references}
\end{document}